\documentclass[a4paper,11pt]{article}

\usepackage{jinstpub} 
\usepackage{booktabs}
\usepackage{lineno}
\usepackage[flushleft]{threeparttable}

\title{\boldmath Parametrizing the reconstruction performance of Super-Kamiokande in the Sub-GeV to TeV neutrino energy range}

\author{C. Jes\'us-Valls}
\affiliation{European Organization for Nuclear Research (CERN), 1211 Geneva 23, Switzerland}
\emailAdd{cesar.jesus@cern.ch}

\abstract{Super-Kamiokande is a paramount detector for studying atmospheric, astrophysical and accelerator neutrino physics. This work extracts and characterizes the neutrino reconstruction performance of Super-Kamiokande using the public data release from its latest atmospheric neutrino analysis. Energy and zenith angle reconstruction performances are derived and modeled for neutrinos spanning sub-GeV to TeV energies across different event samples and sample-specific and sample-aggregated performances are provided. These metrics enable realistic detector response modeling in phenomenological studies, establish benchmarks for evaluating alternative reconstruction algorithms, facilitate quantitative performance comparisons with other experiments, and provide a baseline performance model for the future Hyper-Kamiokande detector. This study represents the first publicly available comprehensive model to describe neutrino reconstruction in Super-Kamiokande in the Sub-GeV to TeV energies.}

\begin{document}
\maketitle
\flushbottom

\section{Introduction}
\subsection{The Super-Kamiokande Detector}
Super-Kamiokande (SK)~\cite{Super-Kamiokande:2002weg, Suzuki:2019jby} is a cylindrical water-Cherenkov detector located underground in the Kamioka mine, Japan. The detector consists of a stainless steel tank 39.6 meters in diameter and 42.2 meters in height, containing 50 ktons of ultra-pure water. The detector is divided into two optically isolated regions: an inner detector (ID) containing 32 ktons of water viewed by 11,000 inward-facing 50-cm diameter photomultiplier tubes (PMTs) that provide 40\% photocathode coverage of the inner surface, and an outer detector (OD) consisting of an approximately 2-meter water layer viewed by nearly 2,000 outward-facing 20-cm diameter PMTs equipped with wavelength-shifting plates to enhance light collection. The PMTs are mounted on a stainless steel support structure that separates the two regions. The detector's fiducial volume, defined as the region within the ID at a sufficient distance from the PMT plane to minimize edge effects and external backgrounds, ranges from 22.5 to 27.5 ktons depending on the specific analysis requirements. SK has been collecting data since 1996 and remains operational as one of the most important neutrino detectors worldwide.

\subsection{Neutrino Physics Program}
Super-Kamiokande has been instrumental in establishing neutrino oscillations through its observation of zenith-angle-dependent deficits in atmospheric muon neutrinos~\cite{Super-Kamiokande:1998kpq} and, together with SNO, the confirmation of solar neutrino deficits~\cite{Super-Kamiokande:2001ljr, SNO:2002tuh}. The detector has also played a pioneering role in accelerator neutrino oscillation studies, first with K2K~\cite{K2K:2002icj} and currently with T2K~\cite{T2K:2011qtm, T2K:2019bcf}.
Nearly three decades of data collection at Super-Kamiokande have enabled precision measurements that now underpin fundamental aspects of neutrino physics, including the standard solar model~\cite{Asplund:2009fu}, the atmospheric neutrino flux predictions~\cite{Super-Kamiokande:2015qek}, and the phenomenological framework for neutrino oscillations~\cite{Esteban:2024eli}. SK continues to collect data and improve its measurements through both analysis refinements and hardware upgrades. Notable among these upgrades is the addition of gadolinium in 2018~\cite{Super-Kamiokande:2021the, Super-Kamiokande:2025rgm}, which has been crucial for detecting the first hints of the diffuse supernova neutrino background~\cite{Santos:2025plx, Beacom:2003nk}.

\subsection{Reconstruction Performances}
The wide energy range of neutrinos studied at Super-Kamiokande (SK) has necessitated the development of independent reconstruction strategies and algorithms for different physics programs. The reconstruction performance for low-energy neutrinos (tens of MeV), which is crucial for solar and supernova neutrino studies, employs the Bonsai algorithm~\cite{Smy:2007maa} and is well documented in Refs.~\cite{Super-Kamiokande:2010tar, Super-Kamiokande:2023jbt}. In contrast, comprehensive studies about the detector reconstruction performances at higher energies, essential for accelerator, atmospheric, and astrophysical neutrino studies, can not be found in the scientific literature.

Initially, SK used the APfit algorithm for both atmospheric and accelerator neutrino measurements. Subsequently, the T2K collaboration developed FitQun, a likelihood-based reconstruction method derived from MiniBooNE's reconstruction framework~\cite{Patterson:2009ki}. The SK collaboration later adopted FitQun for atmospheric studies as well. Performance metrics comparing APfit and FitQun are available in Ref.~\cite{Super-Kamiokande:2019gzr}, demonstrating FitQun's superior performance. While these metrics provide information about the detector's ability to reconstruct final-state charged lepton kinematics, their utility for many physics studies remains limited.

From a physics perspective, metrics for neutrino energy and direction resolution are often more valuable, as these variables directly relate to measurements of interest. However, obtaining such metrics presents several challenges. Neutrino interaction channels are energy-dependent, resulting in markedly different event topologies between sub-GeV and multi-GeV interactions in Super-Kamiokande. Consequently, SK data are typically divided into separate samples to address each energy regime independently, with reconstruction performance varying significantly across samples. The latest SK atmospheric studies utilized 29 such samples~\cite{Super-Kamiokande:2023ahc}, reflecting the experiment's long history and rich dataset. This diversity and complexity, combined with ongoing efforts to incorporate additional samples, may explain why comprehensive metrics for neutrino energy and angular reconstruction at SK remain unavailable in an accessible, clear, and readily usable format.

This work addresses this gap by providing benchmark numbers for neutrino reconstruction performance that the community can use as inputs for future physics studies. Since Hyper-Kamiokande (HK)~\cite{Hyper-Kamiokande:2018ofw} plans to use FitQun as its initial reconstruction algorithm and shares detector similarities with SK, the metrics presented here can serve as approximate performance estimates for HK. These estimates can be refined using the same methodology once comparable HK data become publicly available.

\section{Methodology}
In conjunction with its latest atmospheric neutrino analysis~\cite{Super-Kamiokande:2023ahc}, the Super-Kamiokande collaboration released a public dataset containing bin-level information for all analysis samples~\cite{SuperKamiokande2023}. This dataset includes the true neutrino energy RMS ($E_{\text{RMS}}$) and mean ($E_{\text{AVG}}$), as well as the true neutrino zenith angle RMS ($c^{\text{RMS}}_{\theta}$) and mean ($c^{\text{AVG}}_{\theta}$), where $c_{\theta}$ denotes the cosine of the zenith angle $\theta$.

In this study that information is used to develop parametric models both for individual samples and for aggregated samples with similar characteristics. The procedure prioritizes finding a generic parametrization that presents the results in a form that is simple and directly applicable for the community. The fitted functions predict the neutrino energy resolution ($E_{\text{RMS}}/E_{\text{AVG}}$) and the neutrino direction angular spread ($c^{\text{RMS}}_{\theta}$) as functions of the neutrino energy.

\subsection{Sample Characteristics and Classification}
\begin{table}[htbp]
\centering
\small
\caption{List of neutrino samples and the flavor used to parametrize reconstruction performance for each.}
\label{tab:neutrino_samples}
\begin{tabular}{c|l|c}
\toprule
ID & Sample Name & Neutrino Type \\
\midrule
 1 & Sub-GeV $\nu_e$-like & $\nu_e$ \\
 2 & Sub-GeV $\bar{\nu}_e$-like 0 n & $\bar{\nu}_e$ \\
 3 & Sub-GeV $\bar{\nu}_e$-like 1 n & $\bar{\nu}_e$ \\
 4 & Sub-GeV $\nu_\mu$-like & $\nu_\mu$ \\
 5 & Sub-GeV $\bar{\nu}_\mu$-like & $\bar{\nu}_\mu$ \\
 6 & Multi-GeV $\nu_e$-like & $\nu_e$ \\
 7 & Multi-GeV $\bar{\nu}_e$-like 0 n & $\bar{\nu}_e$ \\
 8 & Multi-GeV $\bar{\nu}_e$-like 1 n & $\bar{\nu}_e$ \\
 9 & Multi-GeV $\nu_\mu$-like & $\nu_\mu$ \\
10 & Multi-GeV $\bar{\nu}_\mu$-like & $\bar{\nu}_\mu$ \\
11 & Multi-GeV Multi-Ring $\nu_e$-like & $\nu_e$ \\
12 & Multi-GeV Multi-Ring $\bar{\nu}_e$-like & $\bar{\nu}_e$ \\
13 & Multi-GeV Multi-Ring $\mu$-like & $\nu_\mu$ \\
14 & PC Stopping & $\nu_\mu$ \\
15 & PC Through-going & $\nu_\mu$ \\
16 & Up-$\mu$ Stopping & $\nu_\mu$ \\
17 & Up-$\mu$ Non-Showering & $\nu_\mu$ \\
18 & Up-$\mu$ Showering & $\nu_\mu$ \\
\bottomrule
\end{tabular}
\end{table}
The Super-Kamiokande atmospheric neutrino data are categorized into distinct samples based on event topology, energy, and particle identification to enhance sensitivity to different oscillation signals.  Full details are available in Ref.~\cite{Super-Kamiokande:2023ahc}, we provide a summary below. The samples are broadly divided into three main categories: fully-contained (FC), partially-contained (PC), and upward-going muons (Up-$\mu$).

\subsubsection{Fully-Contained Events}
Fully-contained events have their interaction vertex within the inner detector and minimal outer detector activity. These events span energies from 100~MeV to 100~GeV and contain both charged-current (CC) and neutral-current (NC) interactions. FC events are first classified by ring topology:\\

\noindent \textbf{Single-Ring Events} are separated by particle identification (e-like or $\mu$-like) based on Cherenkov ring patterns. Electrons and photons produce diffuse rings due to electromagnetic showers, while muons create sharp-edged rings. Events are further divided by visible energy into sub-GeV ($E_{\text{vis}} < 1330$~MeV) and multi-GeV ($E_{\text{vis}} > 1330$~MeV) categories. The number of decay electrons provides additional separation to enhance neutrino/anti-neutrino purity. For SK-IV and SK-V periods, neutron tagging information (with 26\% efficiency) provides enhanced separation between neutrinos and anti-neutrinos, as anti-neutrino interactions produce more neutrons on average through proton-to-neutron conversions and higher hadronic energy transfer.\\

\noindent \textbf{Multi-Ring Events} at multi-GeV energies are classified using a boosted decision tree (BDT) into $\nu_e$-like, $\bar{\nu}_e$-like, $\mu$-like, and ``other'' (primarily NC) categories. The BDT uses seven variables including ring count, decay electrons, particle ID, energy fractions, and transverse momentum.

\subsubsection{Partially-Contained Events}
Partially-contained events have vertices within the inner detector but produce muons that exit into the outer detector, with typical energies between 1~GeV and 1~TeV. These are nearly all $\nu_\mu$ CC interactions and are classified as:
\begin{itemize}\setlength{\itemsep}{0pt}
\item \textbf{Stopping}: Muons that stop within the outer detector
\item \textbf{Through-going}: Muons that completely exit the detector
\end{itemize}

\subsubsection{Upward-Going Muons}
Upward-going muon events originate from neutrino interactions in the rock below the detector or in the outer detector water, reaching energies up to $\sim$10~TeV. Earth shielding eliminates cosmic ray backgrounds below the horizon. These highest-energy events are classified into three categories:
\begin{itemize}\setlength{\itemsep}{0pt}
\item \textbf{Stopping}: Muons that stop within the inner detector
\item \textbf{Showering}: Exiting muons with charge deposition consistent with radiative losses
\item \textbf{Non-showering}: Exiting muons without significant radiative losses
\end{itemize}

\subsection{Analysis Samples}

For this analysis, we focus on the Super-Kamiokande data from the most recent detector configurations (SK-IV and SK-V), which best represent the current detector performance. From the complete dataset, we exclude three samples\footnote{Those that rely on $\pi^0$ identification and the Multi-GeV Multi Ring ``Other'' sample.} where interactions are dominated by neutral current (NC) events, as these are less relevant for most studies. This results in a total of 18 analysis samples.

Each sample contains different fractions of neutrino flavors, leading to distinct reconstruction performances. Thanks to Super-Kamiokande's excellent lepton flavor identification capabilities, samples maintain high purity for specific neutrino types. We therefore parametrize the reconstruction performance of each sample based on the reconstruction performance for its dominant neutrino flavor. Table~\ref{tab:neutrino_samples} summarizes the 18 samples used in this analysis and indicates the neutrino flavor used to characterize the reconstruction performance for each sample.

\subsection{Sample-by-Sample Fitting}
Sample-by-sample modeling involves a two-stage fitting process.

\subsubsection*{First Fitting Stage}
\begin{itemize}
    \item \textbf{Linear:} For samples with two energy bins per angle bin, a linear parametrization is used: $E_{\text{RMS}}/ E_{\text{AVG}} = a_E + b_E\times E$ and $c^{\text{RMS}}_{\theta} = a_A + b_A\times E$
    \item \textbf{Power-Law:} For samples with more than two energy bins per angle bin, a power-law parametrization is used: $E_{\text{RMS}}/ E_{\text{AVG}} = a_E \times E^{-b_E}$ and $c^{\text{RMS}}_{\theta} = a_A \times E^{-b_A}$
\end{itemize}

\subsubsection*{Second Fitting Stage}
The zenith dependence of the parameters $a_E$ and $a_A$ is fit with a parabolic function: 
\begin{equation}
a_E(\cos\theta_z) = c_0 + c_1 \cos\theta_z + c_2 \cos^2\theta_z
\end{equation}
An analogous parabolic fit is applied to parameters $b_E$, $a_A$, and $b_A$.

\subsection{Fitting Aggregated Samples}
While sample-by-sample fits provide detailed reconstruction benchmarks, phenomenological studies may benefit from a simplified set of functions that reasonably approximate SK performances. 

The aggregated sample fitting approach consists of grouping together samples with similar performance characteristics and performing a single fit across all their bins. We use the same power-law equations introduced previously:
\begin{align}
E_{\text{RMS}}/ E_{\text{AVG}} &= a_E \times E^{-b_E}\\
c^{\text{RMS}}_{\theta} &= a_A \times E^{-b_A}
\end{align}

This strategy simplifies the performance characterization to a total of only 8 equations (4 to model the neutrino energy reconstruction and 4 to model the angular reconstruction), corresponding to natural groupings of the Sub-GeV, Multi-GeV, Multi-GeV Multi-Ring, and Up-$\mu$ samples. Although the energy range of the Partially Contained (PC) samples overlaps with that of the samples listed above, these PC samples are excluded from the combined fits due to their inferior reconstruction performance.

\section{Results}
\begin{figure*}[htbp]
\centering
\includegraphics[width=\textwidth]{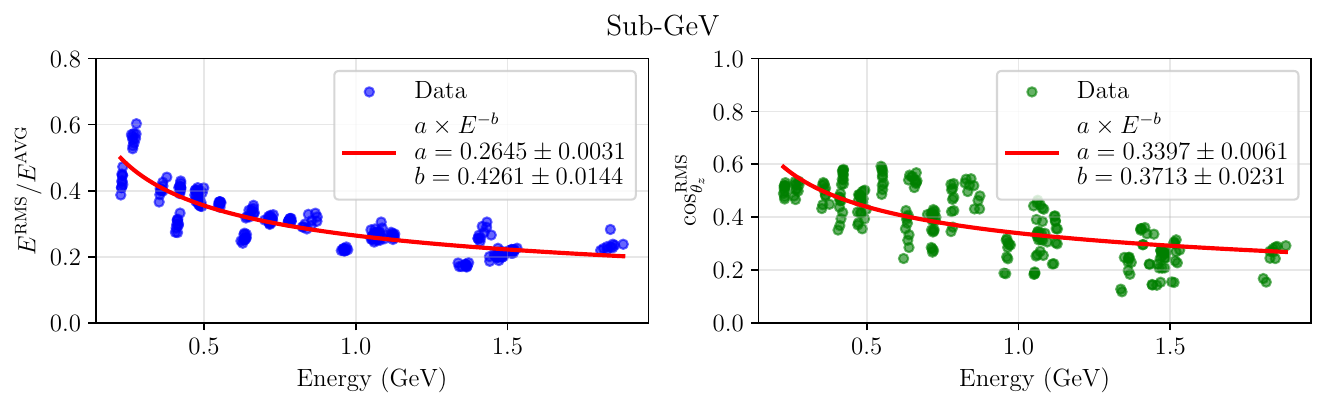}
\includegraphics[width=\textwidth]{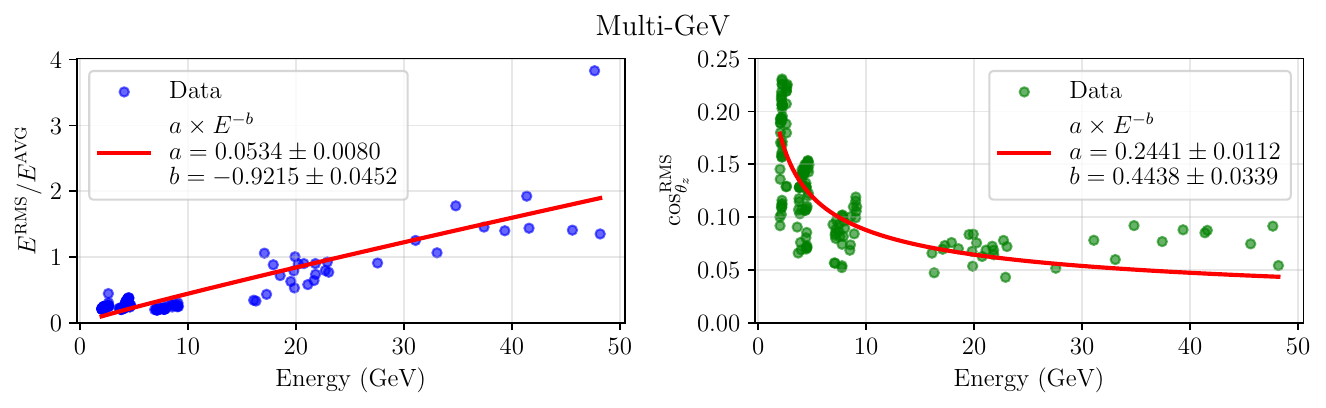}
\includegraphics[width=\textwidth]{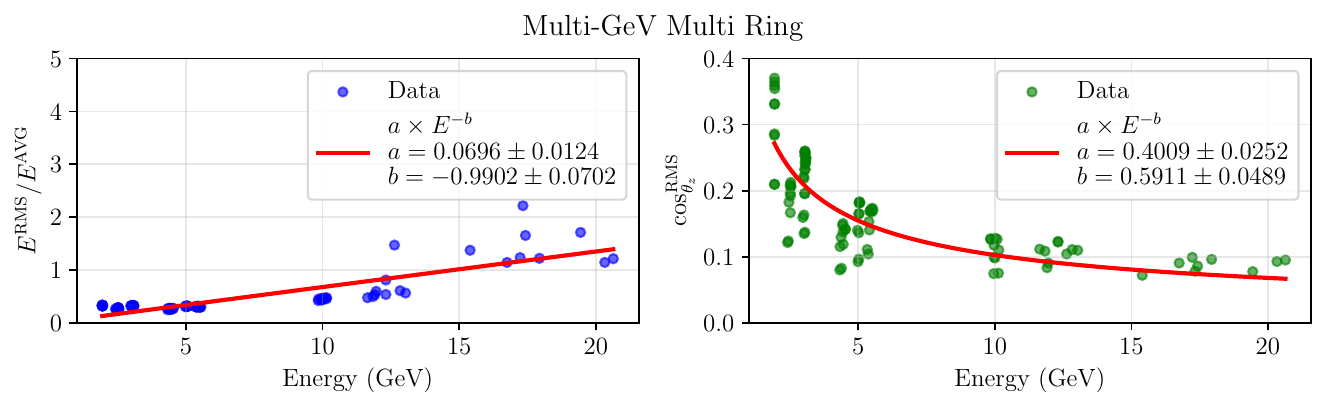}
\includegraphics[width=\textwidth]{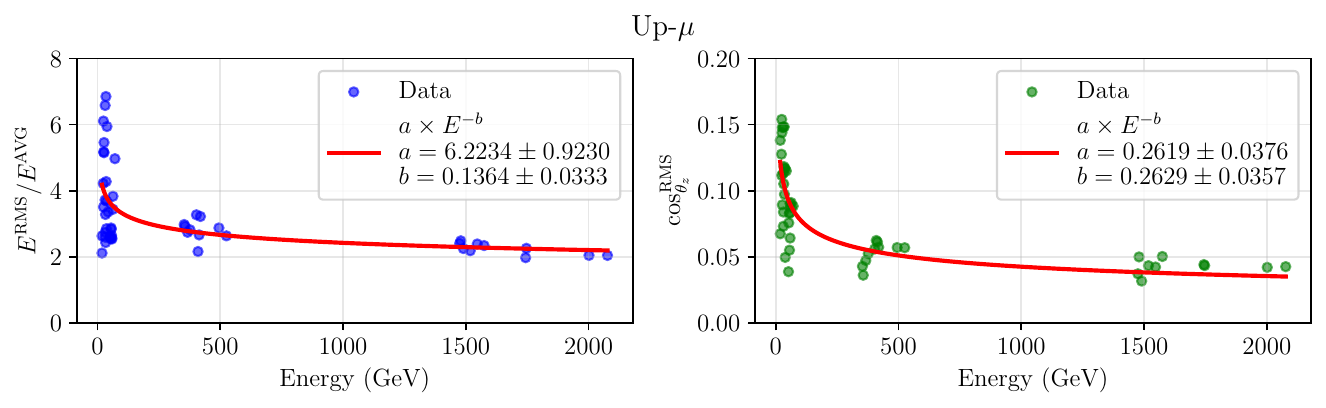}
\caption{Energy and angular reconstruction performances for Super-Kamiokande using fits to groups of samples. From top to bottom, and in accordance to  Tab.~\ref{tab:neutrino_samples}: Sub-GeV samples (IDs: 0-5); Multi-GeV samples (IDs 6-10); Multi-GeV Multi Ring samples (IDs 11-13) and Up-$\mu$ samples IDs (16-18). PC samples are not considered.}
\label{fig:combined_fits}
\end{figure*}

The results of the aggregated sample analysis are presented in Fig.~\ref{fig:combined_fits}. Several physical phenomena affect the reconstruction performance across different energy regimes:

Fermi motion introduces a constant amount of energy smearing that significantly impacts low-energy Sub-GeV events but becomes subdominant as energy increases, resulting in improved energy resolution with increasing energy for Sub-GeV samples. In contrast, for Multi-GeV and Multi-GeV Multi-Ring samples, energy resolution degrades with increasing energy because a substantial fraction of the neutrino energy is transferred to final-state hadrons, which are often undetectable in water Cherenkov detectors. Consequently, optimal energy reconstruction occurs at approximately 1~GeV, with resolutions approaching 20\% of the neutrino energy. For very high energy neutrinos in the Up-{\textmu} samples, interacting outside of the detector, the total light observed in the detector allows to reach resolution similar to 200\% for TeV energies.

The angular reconstruction performance, quantified by the spread in $\cos \theta$, consistently improves with increasing neutrino energy across all samples. This improvement occurs because higher-energy interactions produce more forward-boosted final-state particles, making the overall light directionality a stronger indicator of the incident neutrino direction. The observed pattern shows that the RMS of $\cos\theta$ decreases from approximately 0.3 at 1~GeV to 0.2 for few-GeV events, further improving to 0.1 at approximately 10~GeV, and reaching values as low as 0.05 for TeV-scale energies. Table~\ref{tab:simple_combined_resolution} summarizes the parameters shown in Fig.~\ref{fig:combined_fits}.

While the power-law parameterization effectively captures the general trends, detailed examination of the data reveals sample-specific patterns that warrant a more granular treatment. Tables~\ref{tab:linear_performance} and \ref{tab:power_law_performance} present the results of the sample-by-sample analysis for linearly parameterized and power-law fitted samples, respectively. The corresponding fits used to derive these parameters are provided in Appendix~\ref{sec:appendix_figs}.

\begin{table*}[htbp]
\centering
\small
\caption{Parametrization for aggregated samples: $E^{\small{\textup{RMS}}}/E^{\small{\textup{AVG}}} = a_E \times E^{-b_E}$; $c^{\small{\textup{RMS}}}_{\theta_z} = a_A \times E^{-b_A}$}
\label{tab:simple_combined_resolution}
\hspace*{\fill}
\begin{tabular}{l|@{\hspace{3pt}}c@{\hspace{3pt}}|@{\hspace{3pt}}c@{\hspace{3pt}}|@{\hspace{3pt}}c@{\hspace{3pt}}|@{\hspace{3pt}}c@{\hspace{3pt}}|c}
\toprule
& \multicolumn{2}{@{\hspace{3pt}}c@{\hspace{3pt}}|@{\hspace{3pt}}}{$E^{\small{\textup{RMS}}}/E^{\small{\textup{AVG}}}$} & \multicolumn{2}{@{\hspace{3pt}}c@{\hspace{3pt}}|}{$c^{\small{\textup{RMS}}}_{\theta_z}$} & \\
Sample Group & $a_E$ & $b_E$ & $a_A$ & $b_A$ & $E_{\text{Range}}$ [GeV] \\
\midrule\\
Sub-GeV & 0.26 $\pm$ 0.00 & 0.43 $\pm$ 0.01 & 0.34 $\pm$ 0.01 & 0.37 $\pm$ 0.02 & [0.2, 1.9] \\
Multi-GeV & 0.05 $\pm$ 0.01 & -0.92 $\pm$ 0.05 & 0.24 $\pm$ 0.01 & 0.44 $\pm$ 0.03 & [2.0, 48] \\
Multi-Ring & 0.07 $\pm$ 0.01 & -0.99 $\pm$ 0.07 & 0.40 $\pm$ 0.03 & 0.59 $\pm$ 0.05 & [1.9, 21] \\
Up-$\mu$ & 6.22 $\pm$ 0.92 & 0.14 $\pm$ 0.03 & 0.26 $\pm$ 0.04 & 0.26 $\pm$ 0.04 & [18.4, 2077] \\
\bottomrule
\end{tabular}
\hspace*{\fill}
\end{table*}

\begin{table*}[htbp]
\centering
\small
\caption{Linear Parametrization: $E^{\small{\textup{RMS}}}/E^{\small{\textup{AVG}}} = a_E + b_E \times E$; $c^{\small{\textup{RMS}}}_{\theta_z} = a_A + b_A \times E$ (coefficients $\times 10$)}
\label{tab:linear_performance}
\hspace*{\fill}
\begin{tabular}{l|@{\hspace{3pt}}r@{\hspace{3pt}}r@{\hspace{3pt}}r@{\hspace{3pt}}|@{\hspace{3pt}}r@{\hspace{3pt}}r@{\hspace{3pt}}r@{\hspace{3pt}}|@{\hspace{3pt}}r@{\hspace{3pt}}r@{\hspace{3pt}}r@{\hspace{3pt}}|@{\hspace{3pt}}r@{\hspace{3pt}}r@{\hspace{3pt}}r@{\hspace{3pt}}|c}
\toprule
& \multicolumn{6}{@{\hspace{3pt}}c@{\hspace{3pt}}|@{\hspace{3pt}}}{$E^{\small{\textup{RMS}}}/E^{\small{\textup{AVG}}}$} & \multicolumn{6}{@{\hspace{3pt}}c@{\hspace{3pt}}|}{$c^{\small{\textup{RMS}}}_{\theta_z}$} & \\
\midrule
& \multicolumn{3}{@{\hspace{3pt}}c@{\hspace{3pt}}|@{\hspace{3pt}}}{$a_E$} & \multicolumn{3}{@{\hspace{3pt}}c@{\hspace{3pt}}|@{\hspace{3pt}}}{$b_E$} & \multicolumn{3}{@{\hspace{3pt}}c@{\hspace{3pt}}|@{\hspace{3pt}}}{$a_A$} & \multicolumn{3}{@{\hspace{3pt}}c@{\hspace{3pt}}|}{$b_A$} & \\
ID & $c_0$ & $c_1$ & $c_2$ & $c_0$ & $c_1$ & $c_2$ & $c_0$ & $c_1$ & $c_2$ & $c_0$ & $c_1$ & $c_2$ & $E_{Range}$ \\
\midrule
9 & -0.36 & 0.10 & 1.52 & 0.20 & -0.03 & 0.41 & -1.95 & 0.02 & 3.28 & 0.24 & -0.01 & -0.40 & [2, 5] \\
10 & -0.54 & -0.12 & 1.44 & 0.25 & 0.04 & 0.41 & -1.87 & -0.01 & 3.04 & 0.25 & -0.00 & -0.36 & [2, 5] \\
14 & 1.30 & 0.51 & 0.09 & 0.00 & -0.22 & 1.72 & -1.80 & 0.05 & 3.05 & 0.10 & -0.01 & -0.12 & [2, 14] \\
\bottomrule
\end{tabular}
\hspace*{\fill}
\begin{tablenotes}
\small
\item \textbf{Zenith dependence:} $a_E = c_0 + c_1 \cos\theta_z + c_2 \cos^2\theta_z$ (analogously for $b_E$, $a_A$, $b_A$).
\item \textbf{Range:} Energy range of bins used for fitting each sample in GeV.
\end{tablenotes}
\end{table*}

\begin{table*}[htbp]
\centering
\small
\caption{Power-Law Parametrization: $E^{\small{\textup{RMS}}}/E^{\small{\textup{AVG}}} = a_E \times E^{-b_E}$; $c^{\small{\textup{RMS}}}_{\theta_z} = a_A \times E^{-b_A}$ (coefficients $\times 10$)}
\label{tab:power_law_performance}
\hspace*{\fill}
\begin{tabular}{l|@{\hspace{3pt}}r@{\hspace{3pt}}r@{\hspace{3pt}}r@{\hspace{3pt}}|@{\hspace{3pt}}r@{\hspace{3pt}}r@{\hspace{3pt}}r@{\hspace{3pt}}|@{\hspace{3pt}}r@{\hspace{3pt}}r@{\hspace{3pt}}r@{\hspace{3pt}}|@{\hspace{3pt}}r@{\hspace{3pt}}r@{\hspace{3pt}}r@{\hspace{3pt}}|c}
\toprule
& \multicolumn{6}{@{\hspace{3pt}}c@{\hspace{3pt}}|@{\hspace{3pt}}}{$E^{\small{\textup{RMS}}}/E^{\small{\textup{AVG}}}$} & \multicolumn{6}{@{\hspace{3pt}}c@{\hspace{3pt}}|}{$c^{\small{\textup{RMS}}}_{\theta_z}$} & \\
\midrule
& \multicolumn{3}{@{\hspace{3pt}}c@{\hspace{3pt}}|@{\hspace{3pt}}}{$a_E$} & \multicolumn{3}{@{\hspace{3pt}}c@{\hspace{3pt}}|@{\hspace{3pt}}}{$b_E$} & \multicolumn{3}{@{\hspace{3pt}}c@{\hspace{3pt}}|@{\hspace{3pt}}}{$a_A$} & \multicolumn{3}{@{\hspace{3pt}}c@{\hspace{3pt}}|}{$b_A$} & \\
ID & $c_0$ & $c_1$ & $c_2$ & $c_0$ & $c_1$ & $c_2$ & $c_0$ & $c_1$ & $c_2$ & $c_0$ & $c_1$ & $c_2$ & $E_{Range}$ \\
\midrule
1 & 0.09 & -0.01 & 2.88 & 1.68 & -0.00 & 2.58 & -1.29 & 0.02 & 4.62 & 5.42 & 0.17 & 5.49 & [1, 2] \\
2 & 0.02 & 0.06 & 2.05 & -0.48 & -0.16 & 5.07 & -1.62 & 0.07 & 3.29 & 3.49 & 0.03 & 3.37 & [0, 1] \\
3 & 0.05 & -0.00 & 2.59 & -0.66 & 0.10 & 6.01 & -1.59 & 0.00 & 3.58 & 3.07 & 0.06 & 3.31 & [0, 1] \\
4 & 0.01 & 0.00 & 2.73 & -0.14 & -0.20 & 4.76 & -1.92 & 0.04 & 4.34 & 4.98 & -0.03 & 4.03 & [0, 2] \\
5 & -0.09 & 0.02 & 2.60 & 0.33 & 0.33 & 4.56 & -1.78 & 0.03 & 3.64 & 4.23 & 0.39 & 3.78 & [0, 1] \\
6 & 0.60 & -0.04 & 0.35 & 4.54 & -0.21 & -10.90 & -1.75 & 0.09 & 3.31 & -0.71 & 0.14 & 4.20 & [3, 48] \\
7 & 0.83 & -0.29 & 0.49 & 5.49 & -2.27 & -9.57 & -1.82 & -0.05 & 2.81 & -1.92 & -0.03 & 5.13 & [2, 23] \\
8 & 0.23 & -0.15 & 0.64 & 1.59 & -1.03 & -8.24 & -1.90 & -0.03 & 3.17 & -1.31 & -0.32 & 5.15 & [2, 28] \\
11 & -0.34 & 0.25 & 1.45 & -4.93 & 2.77 & -6.18 & -2.46 & 0.11 & 4.97 & -1.12 & 0.24 & 5.82 & [3, 15] \\
12 & -0.40 & -0.13 & 0.84 & -4.18 & -1.80 & -9.37 & -1.72 & -0.20 & 3.08 & -1.62 & -0.54 & 4.25 & [2, 31] \\
13 & -0.01 & 0.04 & 2.56 & -0.06 & 0.17 & -2.22 & -2.88 & -0.01 & 5.85 & 0.57 & -0.03 & 6.71 & [2, 10] \\
15 & 0.67 & -1.16 & 5.75 & 0.80 & -0.59 & -3.68 & -1.49 & -0.01 & 2.62 & -0.28 & -0.04 & 1.97 & [4, 58] \\
\bottomrule
\end{tabular}
\hspace*{\fill}
\begin{tablenotes}
\small
\item \textbf{Zenith dependence:} $a_E = c_0 + c_1 \cos\theta_z + c_2 \cos^2\theta_z$ (analogously for $b_E$, $a_A$, $b_A$).
\item \textbf{Range:} Energy range of bins used for fitting each sample in GeV.
\end{tablenotes}
\end{table*}

\section{Conclusions}

This work provides the first publicly available model to characterize Super-Kamiokande's neutrino reconstruction performance across the sub-GeV to TeV energy range. Using the latest atmospheric neutrino dataset, the neutrino energy and angle reconstruction performance have been characterized for 18 distinct event samples. The resulting parametric models, available both sample-by-sample and in aggregated forms, fill an important gap in the literature by providing quantitative benchmarks that enable realistic detector response modeling in phenomenological studies, facilitate performance comparisons with other neutrino experiments and alternative reconstruction methods, and establish baseline reconstruction expectations for the future Hyper-Kamiokande detector.

\bibliographystyle{ieeetr}
\bibliography{biblio.bib}

\vfill
\newpage
\appendix

\section{Sample-by-Sample Fits}
\label{sec:appendix_figs}

\begin{figure*}[htbp]
\centering
\includegraphics[width=0.94\textwidth]{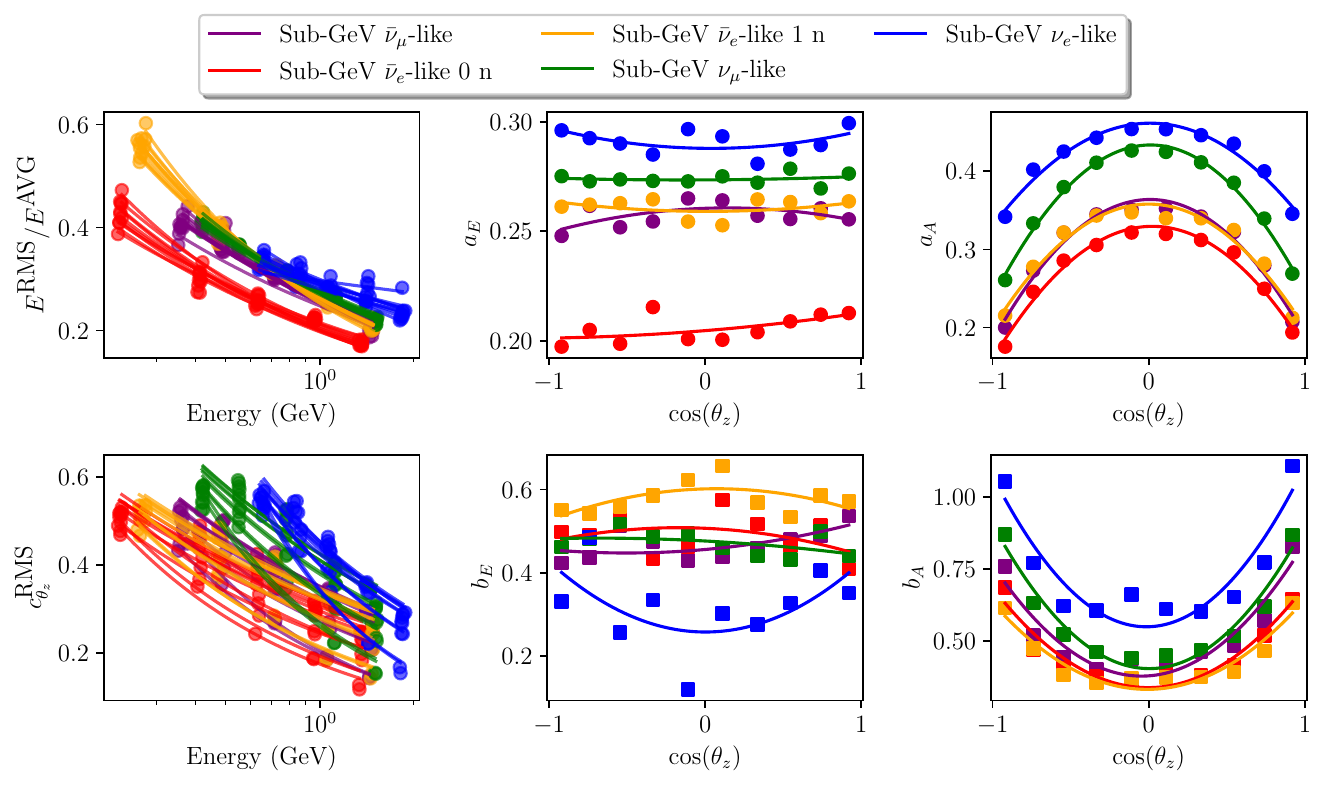}
\caption{First stage fits to Sub-GeV samples (left column) and second stage quadratic fits to the parameters in those fits for the energy (mid column) and angle (right column).}
\label{fig:subgev_resolution_parametrization}
\end{figure*}

\begin{figure*}[htbp]
\centering
\includegraphics[width=0.94\textwidth]{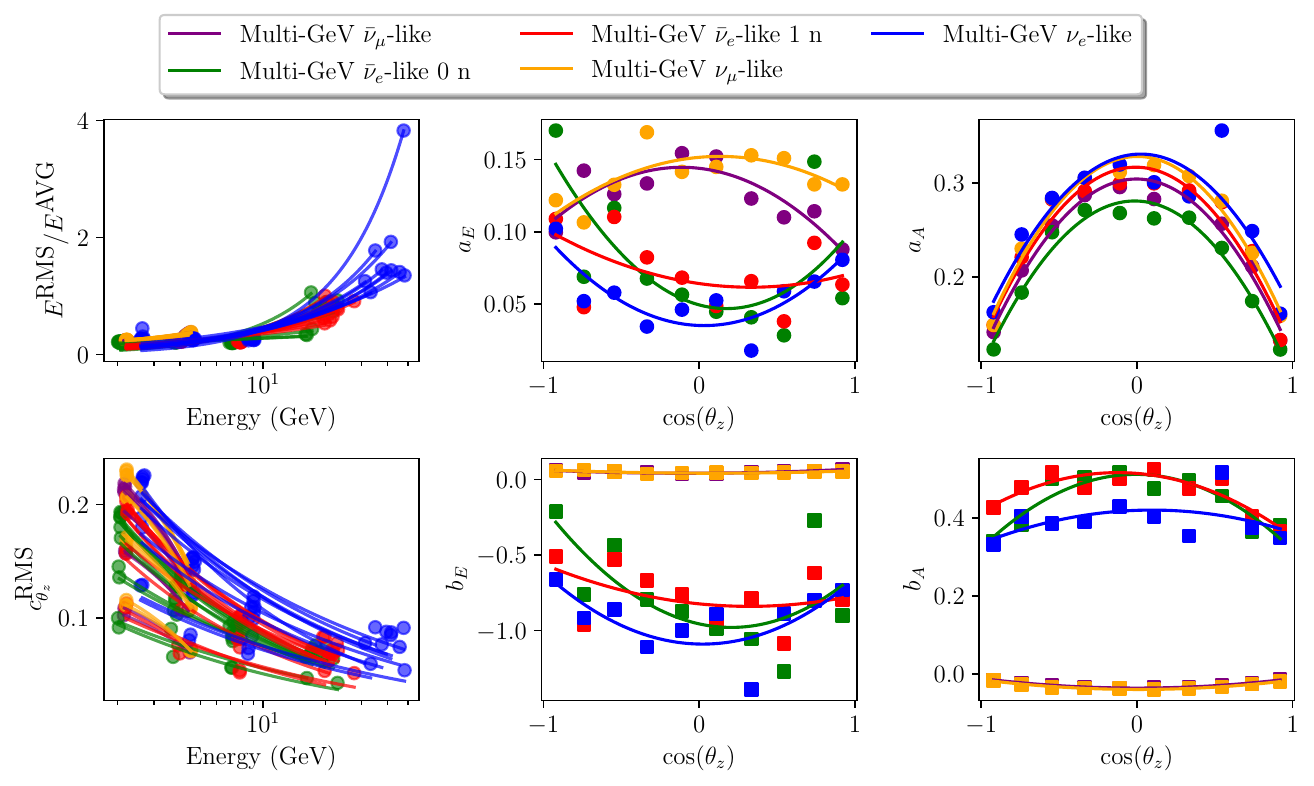}
\caption{First stage fits to Multi-GeV samples (left column) and second stage quadratic fits to the parameters in those fits for the energy (mid column) and angle (right column).}
\label{fig:multigev_resolution_parametrization}
\end{figure*}

\clearpage
\raggedbottom

\begin{figure*}[htbp]
\centering
\includegraphics[width=0.94\textwidth]{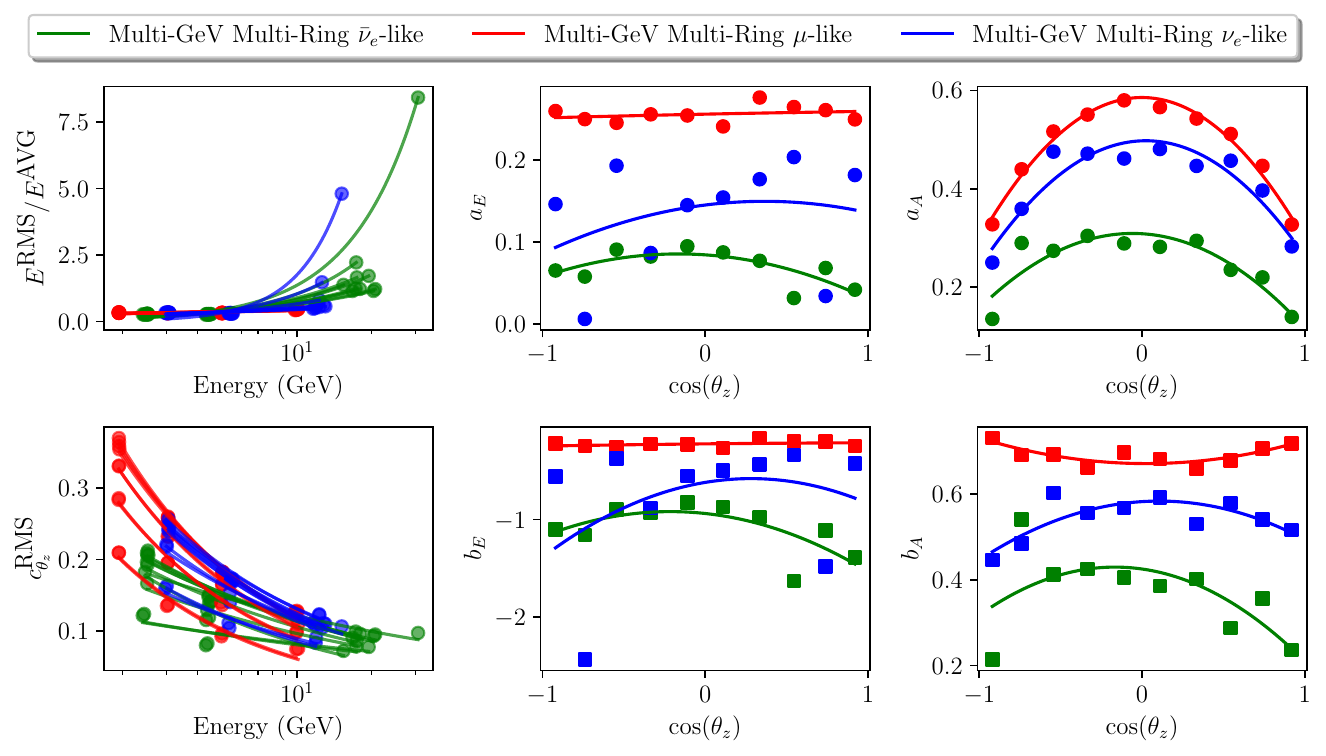}
\caption{First stage fits to Multi-GeV Multi Ring samples (left column) and second stage quadratic fits to the parameters in those fits for the energy (mid column) and angle (right column).}
\label{fig:multiring_resolution_parametrization}
\end{figure*}

\begin{figure*}[htbp]
\centering
\includegraphics[width=0.94\textwidth]{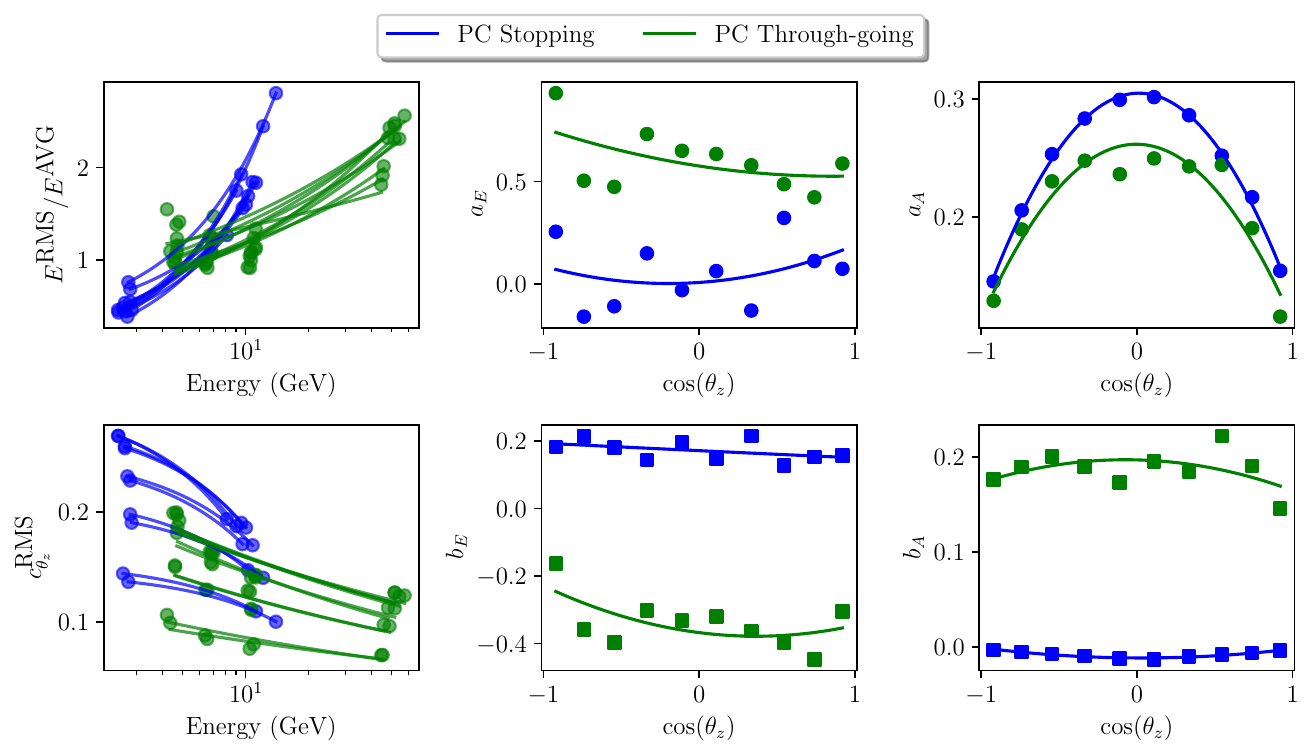}
\caption{First stage fits to Up-{\textmu} samples (left column) and second stage quadratic fits to the parameters in those fits for the energy (mid column) and angle (right column).}
\label{fig:other_resolution_parametrization}
\end{figure*}

\clearpage
\flushbottom

\end{document}